\documentclass[english]{article}
\usepackage[T1]{fontenc}
\usepackage[latin9]{inputenc}
\usepackage{amssymb}
\usepackage{setspace}
\usepackage{esint}

\makeatletter
\newcommand{\lyxaddress}[1]{
\par {\raggedright #1
\vspace{1.4em}
\noindent\par}
}

\makeatother

\usepackage{babel}
\begin{document}

\title{\textbf{Classical non-equilibrium statistical mechanics and an ``open
system dynamics'' perspective on quantum-classical analogy} \thanks{This is a first draft of the manuscript. On a side note, the author
is not familiar with the research fields of classical statistical
mechanics or open classical systems, and this work is a spin-off from
the author's research on quantum dynamics, in part motivated by curiosity
about quantum-classical analogy.}}

\author{Li Yu}
\maketitle

\lyxaddress{\begin{center}
\emph{Department of Physics, Harvard University, Cambridge, MA 02138,
USA}
\par\end{center}}
\begin{abstract}
It is well known that the statistics of closed classical systems evolves
according to the Liouville theorem. Here we study the dynamics of
the marginal statistics of classical systems coupled to external degrees
of freedom, by developing a time-local equation of motion using Green's
functions and a series expansion method. We also compare this equation
of motion with its supposed quantum counterpart, namely the quantum
master equation, which we hope could shed some light on quantum-classical
analogy (QCA) from the perspective of ``open system dynamics''.
We notice an apparent exception to QCA in this case, as the first-order
classical equation of motion derived herein contains a term that does
not appear to have a quantum analogue. We also propose possible ways
of getting around this tension, which may help re-establish QCA (in
first perturbative order). We do not draw a definitive conclusion
about QCA in the context of open system dynamics but hope to provide
a starting point for investigations along this line.
\end{abstract}

\section{Motivations }

\subsection*{Classical non-equilibrium statistical mechanics}

\begin{singlespace}
The probability density of a closed classical system evolves according
to an exact, time-local equation of motion, namely the Liouville theorem.
\cite{tolman book} This readily gives a clear understanding of the
dynamics of a closed classical system's probability density. However,
if a classical system is coupled to other, ``external'' degrees
of freedom, the evolution of its (marginal) probability density does
not generally follow the Liouville theorem, and there is generally
no simplistic description of its dynamics. We call such a system an
``open classical system''. (See also \cite{kong,bhadra,rey,rey 2,hasegawa}
for previous works on open classical systems through different approaches.)

It would be of theoretical interest to have an equation of motion
for the marginal probability density of general open classical systems.
In doing so, one should be able to see the ``properties'' of the
dynamics of an open classical system's marginal density, exact to
every perturbative order. One may gain insights into the general nature
of such dynamics by examining the equation of motion without necessarily
solving it. On the other hand, most real-world physical systems are
coupled to external degrees of freedom. Thus such an equation of motion
might also have pragmatic value as it may be applied to real-world
open classical systems to study the evolution of their (marginal)
statistics.

 Notably, the author has noticed that the early proposal of the ``time-convolutionless
projection operator'' method \cite{shibata 1977,shibata 1979,shibata 1980}
may be applicable to open classical systems as well. If this is the
case, then it should ultimately yield results equivalent to the results
we obtain in this work, as any formally exact theory should be mathematically
equivalent to any other formally exact theory on the same subject.
However, their derivation is different from ours, wherein they use
more constructs such as the projection operator technique and antichronological
time-ordering. \footnote{It also appears that our derivation leads to an additional term as
compared to the results from their derivation, as will be seen later.} In any case, our formalism is developed independently from \cite{shibata 1977,shibata 1979,shibata 1980};
it is mainly inspired by our work on quantum master equation for open
systems \cite{yu} and in part motivated by curiosity about quantum-classical
analogy \cite{dragoman book}. All the steps in our derivation are
naturally motivated and all the intermediate terms are intuitively
defined. Lastly but importantly, our derivation is direct and ``by
construction'', as it does not use any ``inverse'' as ansatz, whereas
the ``inverse'' was invariably invoked in the derivations in \cite{shibata 1977,shibata 1979,shibata 1980}.
\end{singlespace}
\begin{singlespace}

\subsection*{Quantum-classical analogy from an open system dynamics perspective}
\end{singlespace}

\begin{singlespace}
First, it should be cautioned that what we mean by quantum-classical
analogy here is not necessarily the same thing as the quantum-classical
``correspondence principle''. \cite{shankar book,dragoman book}
Rather, we mean, broadly speaking, some kind of resemblence and/or
parallelism between quantum theory and classical theory of physics.
\cite{dragoman book} Moreover, in the author's opinion, the criteria
for what qualifies as an adequate analogy might involve some extent
of subjectivity. 

Quantum-classical analogy has been studied from various perspectives.
\cite{dragoman book} For example, it has been examined from a ``state''
perspective, by looking at the states of quantum systems in the phase
space representation and their classical counterparts. \cite{dragoman book}
It has also been studied from a ``dynamics'' perspective, with an
emphasis on chaos. \cite{gong brumer 1,gong brumer 2} Our work is
also on dynamics. By developing an equation of motion for the marginal
probability density of a general open classical system, we open up
the possibility of formally comparing it with the master equation
for the reduced density matrix of open quantum systems. (See \cite{yu,breuer,breuer book}
for examples of master equations for open quantum systems.) In doing
so, one might look at quantum-classical analogy from a formal, ``reduced
dynamical'' perspective, that is, by looking at the evolution of
a general open system's marginal statistics in both quantum and classical
settings.

We will outline some background thoughts here. First, we choose to
consider the quantum density operator and what may be considered its
classical counterpart - the probability density distribution of classical
system's states, which we shall loosely refer to as ``classical density''
herein. Second, we note that, under the aforementioned consideration,
there is an (obvious) analogy between the equation of motion for the
quantum density (i.e. the von Neumann equation) and the equation of
motion for the classical density (i.e. the Liouville theorem) in the
closed system scenario. Thus, in this case, we may already reasonably
identify a quantum-classical analogy at the level of closed system
dynamics. Third, we may not have a priori reason to assume that the
quantum-classical analogy found at the level of closed system dynamics
would automatically (or necessarily) imply an adequate quantum-classical
analogy at the level of open system dynamics. Therefore, deriving
an equation of motion for a general open classical system's probability
density may facilitate a better examination of quantum-classical analogy
in the open system scenario. This could be of importance/relevance
considering that most systems in the real world, in both quantum and
classical settings, are open systems.
\end{singlespace}
\begin{singlespace}

\section{Derivations}
\end{singlespace}
\begin{singlespace}

\subsection{Basic contructs}
\end{singlespace}
\begin{singlespace}

\subsubsection*{Phase space and probability density }
\end{singlespace}

\begin{singlespace}
Suppose we are dealing with a system of $N$ particles in one dimension.
Let's denote the position and momentum of the $n$-th particle as
$q_{n}$ and $p_{n}$ respectively. Let's denote the collection of
all classical degrees of freedom as $x\equiv\left\{ q_{1},\,p_{1},\,q_{2},\,p_{2},\ldots q_{N},\,p_{N}\right\} \in\mathbb{P}^{2N}$,
where $\mathbb{P}^{2N}$ is the 2N-dimensional phase space. \cite{pathria book}

In classical statistical mechanics, an important quantity is the probability
density (referred to as the density function in \cite{pathria book})
- it is a scalar field on the phase space. At any time t, an instance
of the probability density function $\rho_{t}(x)$ is a state of the
scalar field. Herein we will use the ``vector'' $\overrightarrow{\rho}$
to denote a state of the field \footnote{The state of the scalar field may be thought of as an (infinite-dimensional)
vector. The term ``vector'' here does not necessarily mean a proper
vector that transforms in certain ways under coordinate transformation
in the rigorous mathematical sense. Rather we simply mean a set of
numbers like a column vector.

$\,$} and $\left\{ \rho(x)\right\} $ to denote the components of the vector.
Each vector component $\rho(x)$ is just (the numerical value of)
the probability density at a ``phase point'' $x$ \cite{pathria book,tolman book}.
\end{singlespace}
\begin{singlespace}

\subsubsection*{Liouville's theorem }
\end{singlespace}

\begin{singlespace}
The equation of motion (EOM) for the evolution of probability density
in classical statistical mechanics is given by the Liouville theorem
\cite{pathria book,tolman book}
\begin{equation}
\frac{\partial}{\partial t}\rho_{t}(x)=\left\{ H_{t}(x),\,\rho_{t}(x)\right\} _{x},\label{eq:Liouville eqn}
\end{equation}
where $H_{t}(x)$ is the Hamiltonian and the Poisson bracket on arbitrary
functions $f_{t}(x)$ and $g_{t}(x)$ with respect to variables $x\equiv\left\{ q_{1},\,p_{1},\,q_{2},\,p_{2},\ldots q_{N},\,p_{N}\right\} $
is defined as \cite{tolman book}
\begin{equation}
\left\{ f_{t}(x),\,g_{t}(x)\right\} _{x}\equiv\sum_{k=1}^{N}\left(\frac{\partial f_{t}(x)}{\partial q_{k}}\frac{\partial g_{t}(x)}{\partial p_{k}}-\frac{\partial f_{t}(x)}{\partial p_{k}}\frac{\partial g_{t}(x)}{\partial q_{k}}\right).
\end{equation}

Note that this EOM is of first order in time, which means the values
of $\left\{ \rho_{t}(x)\right\} $ for $\forall x\in\mathbb{P}^{2N}$
(equivalently the vector $\overrightarrow{\rho}_{t}$) is properly
a ``state'' of the field. That is, $\left\{ \rho_{t}(x)\right\} $
contains all (statistical) information about the physical system under
consideration and all subsequent evolution of the classical statistics
of the system is completely determined by $\left\{ \rho_{t}(x)\right\} $.
We shall call this the ``statistical state'' of the system to avoid
possible confusion with the ``physical state'' of the system, the
latter of which means the system is in some definitive state charaterized
by variables $x$ as in classical mechanics.
\end{singlespace}
\begin{singlespace}

\subsubsection*{Green's function }
\end{singlespace}

\begin{singlespace}
Because of linearity, we introduce the Green's function \cite{wiki grn ftn,byron book}
$G_{t}(x,y)$ formally, such that the following equality holds for
arbitrary $\rho_{0}(y)$:
\begin{equation}
\rho_{t}(x)=\int dy\,G_{t}(x,y)\,\rho_{0}(y).\label{eq:grn ftn def}
\end{equation}

Physically, this equality may be interpreted in the following way:
The probability (density) $\rho_{t}(x)$ at $x$ at time $t$ comes
from the probability sources $\rho_{0}(y)$ all across phase space
$\forall y\in\mathbb{P}^{2N}$ at time $t=0$ with respective weights
$G_{t}(x,y)$. To be more precise, the initial probability $\rho_{0}(y_{1})$
of the system being at physical state $y_{1}$ contributes to the
final probability $\rho_{t}(x)$ with a weight $G_{t}(x,y_{1})$,
the initial probability $\rho_{0}(y_{2})$ at $y_{2}$ contributes
to the final probability $\rho_{t}(x)$ with a weight $G_{t}(x,y_{2})$,
and so on; and we obtain the full final probability $\rho_{t}(x)$
by adding up contributions from all the initial probabilities $\left\{ \rho_{0}(y)\right\} $
for $\forall y\in\mathbb{P}^{2N}$, each contribution weighed by $G_{t}(x,y)$.
In other words, the larger the Green's function $G_{t}(x,y)$ is for
some $y$, the bigger ``bang for the buck'' the initial probability
$\rho_{0}(y)$ at that particular $y$ has on the final probability
$\rho_{t}(x)$.

Let's now work out the Green's functions. First, plugging Eq.(\ref{eq:grn ftn def})
into the Liouville's equation Eq.(\ref{eq:Liouville eqn}) yields
\begin{equation}
\int dy\,\rho_{0}(y)\,\frac{\partial}{\partial t}G_{t}(x,y)=\int dy\,\rho_{0}(y)\,\left\{ H_{t}(x),\,G_{t}(x,y)\right\} _{x}.\label{eq:eom new}
\end{equation}
Now we know that Eq.(\ref{eq:eom new}) must hold for arbitrary $\rho_{0}(y)$,
we may choose to set $\rho_{0}(y)=\delta(y,Y)$ for some parameter
$Y\in\mathbb{P}^{2N}$, where $\delta(a,b)$ is the multi-variate
delta function. Then we have 
\begin{eqnarray}
\int dy\,\delta(y,Y)\,\frac{\partial}{\partial t}G_{t}(x,y) & = & \int dy\,\delta(y,Y)\,\left\{ H_{t}(x),\,G_{t}(x,y)\right\} _{x},\\
\Rightarrow\;\;\;\;\;\;\;\;\;\;\;\;\;\frac{\partial}{\partial t}G_{t}(x,Y) & = & \left\{ H_{t}(x),\,G_{t}(x,Y)\right\} _{x}.\label{eq:grn eom old}
\end{eqnarray}
Moreover, the parameter $Y$ is arbitrary, therefore Eq.(\ref{eq:grn eom old})
must hold for $\forall Y\in\mathbb{P}^{2N}$. Thus we conclude that
the Green's function $G_{t}(x,y)$ formally introduced in Eq.(\ref{eq:grn ftn def})
obeys the following EOM 
\begin{equation}
\frac{\partial}{\partial t}G_{t}(x,y)=\left\{ H_{t}(x),\,G_{t}(x,y)\right\} _{x},\label{eq:grn eom}
\end{equation}
for $\forall x,y\in\mathbb{P}^{2N}$ and for all $t$.
\end{singlespace}
\begin{singlespace}

\subsection{Perturbative method}
\end{singlespace}
\begin{singlespace}

\subsubsection*{Power series expansion}
\end{singlespace}

\begin{singlespace}
Let's parametrize the Hamiltonian by a prefactor $\lambda$, which
can be thought of as controlling the strength of the Hamiltonian $\lambda H_{t}(x)$.
Then the EOM for the Green's function Eq.(\ref{eq:grn eom}) becomes
\begin{equation}
\frac{\partial}{\partial t}G_{t}(x,y)=\lambda\left\{ H_{t}(x),\,G_{t}(x,y)\right\} _{x}.\label{eq:grn eom new}
\end{equation}

By analogy with the perturbation technique widely used in quantum
mechanics \cite{shankar book,yu,james1,james2}, let's suppose the
solution to this EOM can be written as a power series in the parameter
$\lambda$: 
\begin{equation}
G_{t}(x,y)=\sum_{n=0}^{\infty}\lambda^{n}G_{t,n}(x,y).\label{eq:series expansion}
\end{equation}
Plugging Eq.(\ref{eq:series expansion}) into Eq.(\ref{eq:grn eom new}),
we have 
\begin{equation}
\sum_{n=0}^{\infty}\lambda^{n}\frac{\partial}{\partial t}G_{t,n}(x,y)=\sum_{n=0}^{\infty}\lambda^{n+1}\left\{ H_{t}(x),\,G_{t,n}(x,y)\right\} _{x}.\label{eq:grn eom new new}
\end{equation}
Because $\lambda$ is arbitrary, in order for Eq.(\ref{eq:grn eom new new})
to hold identically, the equality must hold for each of the $n$th
power term of $\lambda$. That is, 
\begin{eqnarray}
\frac{\partial}{\partial t}G_{t,0}(x,y) & = & 0,\;(n=0)\label{eq:iterative 0}\\
\frac{\partial}{\partial t}G_{t,n}(x,y) & = & \left\{ H_{t}(x),\,G_{t,n-1}(x,y)\right\} _{x}.\;(n=1,2,\ldots)\label{eq:iterative n}
\end{eqnarray}

Eq.(\ref{eq:iterative 0}) implies that $G_{t,0}(x,y)$ does not have
explicit dependence on time. To find its precise form (as function
of $x$ and $y$), let's set $\lambda=0$, which implies 
\begin{equation}
G_{t}(x,y)=\sum_{n=0}^{\infty}\lambda^{n}G_{t,n}(x,y)=G_{t,0}(x,y),
\end{equation}
because all terms in the series Eq.(\ref{eq:series expansion}) vanish
except the zeroth order term. In this case, Eq.(\ref{eq:grn ftn def})
implies 
\begin{eqnarray}
\rho_{t}(x) & = & \int dy\,G_{t}(x,y)\,\rho_{0}(y)\nonumber \\
 & = & \int dy\,G_{t,0}(x,y)\,\rho_{0}(y).\label{eq:G0 old}
\end{eqnarray}
On the one hand, $\lambda=0$ also implies 
\begin{eqnarray}
\frac{\partial}{\partial t}\rho_{t}(x) & = & 0,\\
\Rightarrow\;\;\rho_{t}(x) & = & \rho_{0}(x),\label{eq:const rho}
\end{eqnarray}
because of the vanishing Hamiltonian. Comparing Eq.(\ref{eq:const rho})
with Eq.(\ref{eq:G0 old}), we have 
\begin{equation}
\rho_{0}(x)=\int dy\,G_{t,0}(x,y)\,\rho_{0}(y).\label{eq:G0 requirement}
\end{equation}
In order for Eq.(\ref{eq:G0 requirement}) to hold for arbitrary $x\in\mathbb{P}^{2N}$,
we must have 
\begin{equation}
G_{t,0}(x,y)=\delta(x,y).\label{eq:G0}
\end{equation}

Setting $n=1$ in Eq.(\ref{eq:iterative n}) yields 
\begin{eqnarray}
\frac{\partial}{\partial t}G_{t,1}(x,y) & = & \left\{ H_{t}(x),\,G_{t,0}(x,y)\right\} _{x}\nonumber \\
 & = & \left\{ H_{t}(x),\,\delta(x,y)\right\} _{x},
\end{eqnarray}
which implies 
\begin{equation}
G_{t,1}(x,y)=\int_{0}^{t}dt'\left\{ H_{t'}(x),\,\delta(x,y)\right\} _{x}.\label{eq:G1}
\end{equation}

Similarly, higher-order terms $G_{t,n}(x,y)$ for $n=2,3,\ldots$
in Eq.(\ref{eq:series expansion}) can be worked out iteratively using
Eq.(\ref{eq:iterative n}) and will hereafter be treated as known
functions.
\end{singlespace}
\begin{singlespace}

\subsubsection*{Dynamical map}
\end{singlespace}

\begin{singlespace}
Now we are ready to write down the probability density $\rho_{t}(x)$
at arbitrary $x\in\mathbb{P}^{2N}$ at arbitrary time $t$ as a functional
of the initial state $\overrightarrow{\rho}_{0}$ of the probability
density field on phase space.

With Eqs.(\ref{eq:grn ftn def}, \ref{eq:series expansion}), we have
\begin{eqnarray}
\rho_{t}(x) & = & \int dy\,G_{t}(x,y)\,\rho_{0}(y)\nonumber \\
 & = & \sum_{n=0}^{\infty}\lambda^{n}\int dy\,G_{t,n}(x,y)\,\rho_{0}(y)\nonumber \\
 & \equiv & \sum_{n=0}^{\infty}\lambda^{n}L_{t,n}\left[\overrightarrow{\rho}_{0}\right](x),
\end{eqnarray}
where $L_{t,n}\left[\overrightarrow{\rho}\right](x)$ is a functional
of any ``vector'' $\overrightarrow{\rho}$ (i.e. depending on all
values $\left\{ \rho(y)\right\} $ for $\forall y\in\mathbb{P}^{2N}$)
defined as 
\begin{equation}
L_{t,n}\left[\overrightarrow{\rho}\right](x)\equiv\int dy\,G_{t,n}(x,y)\,\rho(y).
\end{equation}

\end{singlespace}
\begin{singlespace}

\subsection{System's statistics}
\end{singlespace}
\begin{singlespace}

\subsubsection*{System-environment bifurcation}
\end{singlespace}

\begin{singlespace}
Suppose we are interested in the statistics of a subset of variables
only, maybe because we are able to observe/measure those variables
only. Let $x\equiv\left\{ q_{1},\,p_{1},\,q_{2},\,p_{2},\ldots q_{N_{S}},\,p_{N_{S}}\right\} $
denote the degrees of freedom we are interested in and call it the
``system''; let $x'\equiv\left\{ q'_{1},\,p'_{1},\,q'_{2},\,p'_{2},\ldots q'_{N_{E}},\,p'_{N_{E}}\right\} $
denote the other degrees of freedom and call it the ``environment''.
This idea is similar to the bifurcation of system and environment
in quantum mechanics. \cite{breuer book}

The full dynamics of system and environment $\left(x,x'\right)\in\mathbb{P}^{2N}$
obeys the Liouville equation, with $N=N_{S}+N_{E}$, and the same
derivation above still applies. It can be shown that the probability
density at time $t$ is 
\begin{eqnarray}
\rho_{t}(x,x') & = & \int dy\int dy'\,G_{t}(x,x';y,y')\,\rho_{0}(y,y')\nonumber \\
 & = & \sum_{n=0}^{\infty}\lambda^{n}\int dy\int dy'\,G_{t,n}(x,x';y,y')\,\rho_{0}(y,y')\nonumber \\
 & \equiv & \sum_{n=0}^{\infty}\lambda^{n}L_{t,n}\left[\overrightarrow{\rho}_{0}\right](x,x'),\label{eq: S+E prob}
\end{eqnarray}
where $L_{t,n}\left[\overrightarrow{\rho}\right](x,x')$ is a functional
of any ``vector'' $\overrightarrow{\rho}$ (i.e. depending on all
values $\left\{ \rho(y,y')\right\} $ for $\forall(y,y')\in\mathbb{P}^{2N}$)
defined as 
\begin{equation}
L_{t,n}\left[\overrightarrow{\rho}\right](x,x')\equiv\int dy\int dy'\,G_{t,n}(x,x';y,y')\,\rho(y,y'),
\end{equation}
and 
\begin{eqnarray}
G_{t,0}(x,x';y,y') & = & \delta(x,x';y,y')\equiv\delta(x,y)\delta(x',y'),\;(n=0)\label{eq:grn ftn SE 0}\\
G_{t,1}(x,x';y,y') & = & \int_{0}^{t}dt'\left\{ H_{t'}(x,x'),\,\delta(x,y)\delta(x',y')\right\} _{(x,x')},\;(n=1)\label{eq:grn ftn SE 1}\\
G_{t,n}(x,x';y,y') & = & \int_{0}^{t}dt'\left\{ H_{t'}(x,x'),\,G_{t,n-1}(x,x';y,y')\right\} _{(x,x')}.\;(n\geqslant2)\nonumber \\
\label{eq:grn ftn SE 2+}
\end{eqnarray}

\end{singlespace}
\begin{singlespace}

\subsubsection*{Initial statistical independence }
\end{singlespace}

\begin{singlespace}
For the purpose of our discussion, let's suppose at time $t=0$ the
joint probability density is a product of the initial probability
density of the system and that of the environment.

This condition can also be motivated/justified on physical grounds.
Suppose the system's degrees of freedom, say the position and momentum
of a billiard ball, does not interact with the environment's degrees
of freedom, say the position and momentum of a second billiard ball,
before $t=0$; that is, ball 1 and ball 2 have never collided by $t=0$.
In this case, though we are not certain about the either billiard
ball's precise physical state and thus can only describe both in probabilistic
terms $\rho_{0}^{S}(x)$ and $\rho_{0}^{E}(x')$ respectively, we
are certain about the fact that the statistics of ball 1 is independent
of the statistics of ball 2, because now that they have never interacted
they could not have influenced one another. In other words, the joint
probability density is just a product of the two probability densities,
\cite{hodges book,mandel book} 
\begin{equation}
\rho_{0}(x,x')=\rho_{0}^{S}(x)\rho_{0}^{E}(x').\label{eq:initial independence}
\end{equation}

\end{singlespace}
\begin{singlespace}

\subsubsection*{Marginal probability }
\end{singlespace}

\begin{singlespace}
We may integrate the joint probability density over some degrees of
freedom to obtain the marginal probability density on other degrees
of freedom. \cite{hodges book} In particular, if we want to know
the statistics of the system without regards to that of the environment,
we may integrate the joint probability density over the environment's
degrees of freedom $x'\equiv\left\{ q'_{1},\,p'_{1},\,q'_{2},\,p'_{2},\ldots q'_{N_{E}},\,p'_{N_{E}}\right\} $
to find the marginal probability density on the system's degrees of
freedom $x\equiv\left\{ q_{1},\,p_{1},\,q_{2},\,p_{2},\ldots q_{N_{S}},\,p_{N_{S}}\right\} $.
Thus, given the joint probability density $\rho_{t}(x,x')$ at any
time $t$, the marginal probability density of the system is \cite{hodges book}
\begin{eqnarray}
\rho_{t}^{S}(x) & \equiv & \int dx'\rho_{t}(x,x')\nonumber \\
 & = & \sum_{n=0}^{\infty}\lambda^{n}\int dx'\,L_{t,n}\left[\overrightarrow{\rho}_{0}\right](x,x')\nonumber \\
 & = & \sum_{n=0}^{\infty}\lambda^{n}\int dx'\int dy\int dy'\,G_{t,n}(x,x';y,y')\,\rho_{0}(y,y')\nonumber \\
 & = & \sum_{n=0}^{\infty}\lambda^{n}\int dx'\int dy\int dy'\,G_{t,n}(x,x';y,y')\,\rho_{0}^{S}(y)\rho_{0}^{E}(y')\nonumber \\
 & \equiv & \sum_{n=0}^{\infty}\lambda^{n}\mathcal{E}_{t,n}\left[\overrightarrow{\rho}_{0}^{S}\right](x),\label{eq:marginal old}
\end{eqnarray}
where we have made use of Eqs.(\ref{eq: S+E prob}, \ref{eq:initial independence})
and introduced the following definition 
\begin{equation}
\mathcal{E}_{t,n}\left[\overrightarrow{\rho}^{S}\right](x)\equiv\int dy\left(\int dx'\int dy'\,G_{t,n}(x,x';y,y')\rho_{0}^{E}(y')\right)\rho^{S}(y),\label{eq:En general}
\end{equation}
which precisely speaking is a functional of any ``vector'' $\overrightarrow{\rho}^{S}$
(i.e. depending on all values $\left\{ \rho^{S}(y)\right\} $ for
$\forall y\in\mathbb{P}^{2N_{S}}$).

The zeroth-order term of series can be worked out explicitly using
Eq.(\ref{eq:grn ftn SE 0}), 
\begin{eqnarray}
\mathcal{E}_{t,0}\left[\overrightarrow{\rho}^{S}\right](x) & \equiv & \int dy\left(\int dx'\int dy'\,G_{t,0}(x,x';y,y')\rho_{0}^{E}(y')\right)\rho^{S}(y)\nonumber \\
 & = & \int dy\left(\int dx'\int dy'\,\delta(x,y)\delta(x',y')\rho_{0}^{E}(y')\right)\rho^{S}(y)\nonumber \\
 & = & \left(\int dy\delta(x,y)\rho^{S}(y)\right)\,\left(\int dx'\int dy'\delta(x',y')\rho_{0}^{E}(y')\right)\nonumber \\
 & = & \rho^{S}(x)\,\int dx'\rho_{0}^{E}(x')\nonumber \\
 & = & \rho^{S}(x);\label{eq:E0}
\end{eqnarray}
its time derivative is 
\begin{eqnarray}
\dot{\mathcal{E}}_{t,0}\left[\overrightarrow{\rho}^{S}\right](x) & = & \frac{\partial}{\partial t}\mathbb{I}\left[\overrightarrow{\rho}^{S}\right](x)\nonumber \\
 & = & 0.\label{eq:E0 dot}
\end{eqnarray}
The first-order term can be worked out explicitly using Eq.(\ref{eq:grn ftn SE 1}),
\begin{eqnarray}
\mathcal{E}_{t,1}\left[\overrightarrow{\rho}^{S}\right](x) & \equiv & \int dy\left(\int dx'\int dy'\,G_{t,1}(x,x';y,y')\rho_{0}^{E}(y')\right)\rho^{S}(y)\nonumber \\
 & = & \int dy\left(\int dx'\int dy'\,\int_{0}^{t}dt'\left\{ H_{t'}(x,x'),\,\delta(x,y)\delta(x',y')\right\} _{(x,x')}\,\rho_{0}^{E}(y')\right)\rho^{S}(y)\nonumber \\
 & = & \int_{0}^{t}dt'\int dy\rho^{S}(y)\,\int dx'\int dy'\rho_{0}^{E}(y')\nonumber \\
 &  & \,\left(\delta(x',y')\left\{ H_{t'}(x,x'),\,\delta(x,y)\right\} _{x}+\delta(x,y)\left\{ H_{t'}(x,x'),\,\delta(x',y')\right\} _{x'}\right)\nonumber \\
 & = & \int_{0}^{t}dt'\,\int dx'\left\{ H_{t'}(x,x'),\,\int dy\rho^{S}(y)\delta(x,y)\right\} _{x}\,\int dy'\rho_{0}^{E}(y')\delta(x',y')\nonumber \\
 &  & +\int_{0}^{t}dt'\,\int dy\rho^{S}(y)\delta(x,y)\,\int dx'\left\{ H_{t'}(x,x'),\,\int dy'\rho_{0}^{E}(y')\delta(x',y')\right\} _{x'}\nonumber \\
 & = & \int_{0}^{t}dt'\,\int dx'\left\{ H_{t'}(x,x'),\,\rho^{S}(x)\right\} _{x}\,\rho_{0}^{E}(x')\nonumber \\
 &  & +\int_{0}^{t}dt'\,\rho^{S}(x)\,\int dx'\left\{ H_{t'}(x,x'),\,\rho_{0}^{E}(x')\right\} _{x'}\nonumber \\
 & = & \int_{0}^{t}dt'\,\left\{ \left(\int dx'\rho_{0}^{E}(x')H_{t'}(x,x')\right),\,\rho^{S}(x)\right\} _{x}\nonumber \\
 &  & +\int_{0}^{t}dt'\,\left(\int dx'\left\{ H_{t'}(x,x'),\,\rho_{0}^{E}(x')\right\} _{x'}\right)\,\rho^{S}(x);\label{eq:E1}
\end{eqnarray}
its time derivative is 
\begin{eqnarray}
\mathcal{\dot{E}}_{t,1}\left[\overrightarrow{\rho}^{S}\right](x) & = & \left\{ \left(\int dx'\rho_{0}^{E}(x')H_{t}(x,x')\right),\,\rho^{S}(x)\right\} _{x}\nonumber \\
 &  & +\left(\int dx'\left\{ H_{t}(x,x'),\,\rho_{0}^{E}(x')\right\} _{x'}\right)\,\rho^{S}(x).\label{eq:E1 dot}
\end{eqnarray}
Higher-order terms $\mathcal{E}_{t,n}\left[\overrightarrow{\rho}^{S}\right](x)$
for $n\geqslant2$ can as well be worked out using the higher-order
Green's functions $G_{t,n}(x,x';y,y')$ in Eq.(\ref{eq:grn ftn SE 2+}).

In summary, from Eq.(\ref{eq:marginal old}) we have 
\begin{eqnarray}
\rho_{t}^{S}(x) & = & \mathcal{E}_{t,0}\left[\overrightarrow{\rho}_{0}^{S}\right](x)+\sum_{n=1}^{\infty}\lambda^{n}\mathcal{E}_{t,n}\left[\overrightarrow{\rho}_{0}^{S}\right](x)\nonumber \\
 & = & \rho_{0}^{S}(x)+\mathcal{E}_{t}\left[\overrightarrow{\rho}_{0}^{S}\right](x),\label{eq:E def}
\end{eqnarray}
where 
\begin{equation}
\mathcal{E}_{t}\left[\overrightarrow{\rho}^{S}\right](x)\equiv\sum_{n=1}^{\infty}\lambda^{n}\mathcal{E}_{t,n}\left[\overrightarrow{\rho}^{S}\right](x),\label{eq:E def series}
\end{equation}
with $\mathcal{E}_{t,n}\left[\overrightarrow{\rho}^{S}\right](x)$
defined as in Eq.(\ref{eq:En general}).
\end{singlespace}
\begin{singlespace}

\subsection{Equation of motion}
\end{singlespace}
\begin{singlespace}

\subsubsection*{The $Y_{N,t}$ map}
\end{singlespace}

\begin{singlespace}
Following the approach in \cite{yu}, let's define a linear map that
will be central to our derivation/construction: 
\begin{equation}
Y_{N,t}\left[\overrightarrow{\rho}^{S}\right](x)\equiv\sum_{n=0}^{N}(-1)^{n}\mathfrak{\mathcal{E}}_{t}^{(n)}\left[\overrightarrow{\rho}^{S}\right](x),\label{eq:Y map}
\end{equation}
where $\mathfrak{\mathcal{E}}_{t}^{(n)}\left[\overrightarrow{\rho}^{S}\right](x)\equiv\mathfrak{\mathcal{E}}_{t}\left(\mathfrak{\mathcal{E}}_{t}\left(...\mathfrak{\mathcal{E}}_{t}\left[\overrightarrow{\rho}^{S}\right]\right)\right)(x)$
is a composition of $n$ $\mathfrak{\mathcal{E}}_{t}$ maps.

Note that Eq.(\ref{eq:E def series}) implies $\mathcal{E}_{t}\left[\overrightarrow{\rho}^{S}\right](x)\sim\mathcal{O}\left(\lambda\right)$
and thus $\mathcal{E}_{t}^{(n)}\left[\overrightarrow{\rho}^{S}\right](x)\sim\mathcal{O}\left(\lambda^{n}\right)$.
\end{singlespace}
\begin{singlespace}

\subsubsection*{First-order equation of motion}
\end{singlespace}

\begin{singlespace}
Following the same reasoning as in \cite{yu}, if we want to obtain
an $M$th-order approximate equation of motion, we can choose $N=M-1$
in Eq.(\ref{eq:Y map}), so that the neglected terms are of the order
$\mathcal{O}\left(\lambda^{N+2}\right)\sim\mathcal{O}\left(\lambda^{M+1}\right)$.
Thus, to obtain a first-order equation of motion, let's apply the
Y map Eq.(\ref{eq:Y map}) to $\overrightarrow{\rho}_{t}^{S}$ with
$N=M-1=0$ , which yields 
\begin{eqnarray}
Y_{0,t}\left[\overrightarrow{\rho}_{t}^{S}\right](x) & = & (-1)^{0}\mathbb{I}\left[\overrightarrow{\rho}_{t}^{S}\right](x)\nonumber \\
 & = & \rho_{t}^{S}(x)\nonumber \\
 & = & \rho_{0}^{S}(x)+\mathcal{E}_{t}\left[\overrightarrow{\rho}_{0}^{S}\right](x),\label{eq:Y0 map}
\end{eqnarray}
where in the last equality we have made use of Eq.(\ref{eq:E def}).
Rearranging terms in Eq.(\ref{eq:Y0 map}), we can express $\rho_{0}^{S}(x)$
in terms of $\rho_{t}^{S}(x)$ up to zeroth order 
\begin{eqnarray}
\rho_{0}^{S}(x) & = & \rho_{t}^{S}(x)-\mathcal{E}_{t}\left[\overrightarrow{\rho}_{0}^{S}\right](x)\nonumber \\
 & = & \rho_{t}^{S}(x)+\mathcal{O}\left(\lambda\right),
\end{eqnarray}
or, in an alternative ``vector'' notation, 
\begin{equation}
\overrightarrow{\rho}_{0}^{S}=\overrightarrow{\rho}_{t}^{S}+\mathcal{O}\left(\lambda\right).\label{eq:inversion 1}
\end{equation}

Now, to derive an equation of motion up to first order, we partial
differentiate both sides of Eq.(\ref{eq:E def}) with respect to time,
which yields 
\begin{eqnarray}
\frac{\partial}{\partial t}\rho_{t}^{S}(x) & = & \frac{\partial}{\partial t}\rho_{0}^{S}(x)+\frac{\partial}{\partial t}\mathcal{E}_{t}\left[\overrightarrow{\rho}_{0}^{S}\right](x)\nonumber \\
 & = & 0+\mathcal{\dot{E}}_{t}\left[\overrightarrow{\rho}_{0}^{S}\right](x)\nonumber \\
 & = & \mathcal{\dot{E}}_{t}\left[\overrightarrow{\rho}_{t}^{S}\right](x)+\mathcal{O}\left(\lambda^{2}\right)\nonumber \\
 & = & \lambda\mathcal{\dot{E}}_{t,1}\left[\overrightarrow{\rho}_{t}^{S}\right](x)+\mathcal{O}\left(\lambda^{2}\right),\label{eq:eom 1 old}
\end{eqnarray}
where in the third equality we have made use of Eq.(\ref{eq:inversion 1})
and in the last equality we have made use of Eq.(\ref{eq:E def series}).
Neglecting $\mathcal{O}\left(\lambda^{2}\right)$ terms in Eq.(\ref{eq:eom 1 old}),
using Eq.(\ref{eq:E1 dot}) for the first term, and setting $\lambda=1$
formally, we arrive at a first-order equation of motion: 
\begin{eqnarray}
\frac{\partial}{\partial t}\rho_{t}^{S}(x) & = & \mathcal{\dot{E}}_{t,1}\left[\overrightarrow{\rho}_{t}^{S}\right](x)\nonumber \\
 & = & \left\{ \left(\int dx'\rho_{0}^{E}(x')H_{t}(x,x')\right),\,\rho_{t}^{S}(x)\right\} _{x}\nonumber \\
 &  & +\left(\int dx'\left\{ H_{t}(x,x'),\,\rho_{0}^{E}(x')\right\} _{x'}\right)\,\rho_{t}^{S}(x).
\end{eqnarray}

Note that in order to obtain an equation of motion up to first order
only, we have not followed the full treatment as we did in \cite{yu},
but have instead used a somewhat short-cut approach here. But it is
in the same spirit as the full treatment in \cite{yu}.
\end{singlespace}
\begin{singlespace}

\subsection{Main results}
\end{singlespace}

\begin{singlespace}
So we have a time-local equation of motion for the probability density
$\rho_{t}^{S}(x)$ on the system's phase space $x\equiv\left\{ q_{1},\,p_{1},\,q_{2},\,p_{2},\ldots q_{N_{S}},\,p_{N_{S}}\right\} \in\mathbb{P}^{2N_{S}}$
to first perturbative order 
\begin{eqnarray}
\frac{\partial}{\partial t}\rho_{t}^{S}(x) & = & \left\{ H_{t}^{eff}(x),\,\rho_{t}^{S}(x)\right\} _{x}+B_{t}(x)\rho_{t}^{S}(x),\label{eq:eom 1st order}
\end{eqnarray}
where the effective Hamiltonian $H_{t}^{eff}(x)$ and the prefactor
$B_{t}(x)$ are 
\begin{eqnarray}
H_{t}^{eff}(x) & \equiv & \int dx'\rho_{0}^{E}(x')H_{t}(x,x'),\label{eq:eom def 1}\\
B_{t}(x) & \equiv & \int dx'\left\{ H_{t}(x,x'),\,\rho_{0}^{E}(x')\right\} _{x'}.\label{eq:eom def 2}
\end{eqnarray}

 Provided that the formalism in \cite{shibata 1977,shibata 1979,shibata 1980}
is in fact applicable to the case of open classical systems, the first-order
results therein, albeit in an abstract form, should have served the
purpose of finding first-order corrections to the classical Liouville
dynamics like we have done here. We notice, however, that even to
first perturbative order, our derivation leads to a term $B_{t}(x)\rho_{t}^{S}(x)$
that doesn't appear to have a counterpart in their result. For example,
the first-order term Eq.(31a) of Ref.\cite{shibata 1980} is (roughly
speaking) some \textquotedbl{}average of the perturbing Liouvillian\textquotedbl{},
which is a prima facie counterpart of the first term in Eq.(\ref{eq:eom 1st order})
only. This issue may be worth further investigation. \footnote{The author speculates that, in regard to open classical systems, it
is possible that their derivation might have implicitly assumed some
condition(s) which might be \textquotedbl{}reasonable\textquotedbl{}
for practical purposes but might not be strictly (necessarily) true,
and as a result might have left out something which could have been
included, whereas our treatment might be (arguably) more careful in
the sense that it makes as few assumptions as possible. However, it
must be stressed that such speculation is mere guesswork, based only
on \textquotedbl{}gut feeling\textquotedbl{} but not grounded on any
sound knowledge or rigorous reasoning, and is meant to guess some
possibility for mere exploratary purpose.}
\end{singlespace}
\begin{singlespace}

\section{Discussions}
\end{singlespace}
\begin{singlespace}

\subsection*{Apparent tension with quantum master equation}
\end{singlespace}

\begin{singlespace}
Quantum master equations are used to describe the evolution of open
quantum systems. The quantum master equation can be considered counterpart
to the equation of motion for marginal classical probability density
as we develop in this paper, just like the von-Neumann equation for
closed quantum systems can be considered counterpart to the Liouville
equation for full classical probability density. Thus it makes sense
to compare the main result here with a quantum master equation.

Formally exact master equations can be found in \cite{yu,breuer,breuer book}.
We will compare the main result herein with the master equation in
\cite{yu}. Our equation of motion to first order is Eqs.(\ref{eq:eom 1st order}-\ref{eq:eom def 2})
above, which is to be compared with the first-order master equation
in \cite{yu}: 
\begin{equation}
\frac{d}{dt}\rho_{t}^{S}=-\frac{i}{\hbar}\,\left[H_{eff}(t),\:\rho_{t}^{S}\right],\label{eq:quantum eom 1st order}
\end{equation}
where 
\begin{eqnarray}
H_{eff}(t) & \equiv & \sum_{n}Tr_{E}\left(\rho_{E0}E_{n}(t)\right)S_{n}(t)\nonumber \\
 & = & \sum_{n}Tr_{E}\left(S_{n}(t)\otimes\rho_{E0}E_{n}(t)\right)\nonumber \\
 & = & Tr_{E}\left(\left(\mathbb{I}_{S}\otimes\rho_{E0}\right)\sum_{n}S_{n}(t)\otimes E_{n}(t)\right)\nonumber \\
 & = & Tr_{E}\left(\left(\mathbb{I}_{S}\otimes\rho_{E0}\right)H_{SE}(t)\right).\label{eq:quantum eom def 1}
\end{eqnarray}

Now, if we were to identify the marginal probability density $\rho_{t}^{S}(x)$
in the classical case with the reduced density operator $\rho_{t}^{S}$
in the quantum case, the classical Poisson bracket $\left\{ A(x),\,B(x)\right\} _{x}$
with the quantum commutator $-\frac{i}{\hbar}\,\left[A,\:B\right]$,
the classical effective Hamiltonian density $H_{t}^{eff}(x)\equiv\int dx'\rho_{0}^{E}(x')H_{t}(x,x')$
with the quantum effective Hamiltonian $H_{eff}(t)\equiv Tr_{E}\left(\left(\mathbb{I}_{S}\otimes\rho_{E0}\right)H_{SE}(t)\right)$,
where in turn the integral over classical environmental degrees of
freedom $\int dx'\left(\ldots\right)$ were compared with the partial
trace over quantum environmental degrees of freedom $Tr_{E}\left(\ldots\right)$,
then we can see that the first-order equation of motion developed
herein Eq.(\ref{eq:eom 1st order}) agrees with its quantum counterpart
Eq.(\ref{eq:quantum eom 1st order}) except having an extra term $B_{t}(x)\rho_{t}^{S}(x)$,
where $B_{t}(x)\equiv\int dx'\left\{ H_{t}(x,x'),\,\rho_{0}^{E}(x')\right\} _{x'}$.
Therefore, at face value, we do not appear to have established QCA
at the subsystem (i.e. open system) level even to the lowest (first)
perturbative order.

Note that this is unlike in the case of the whole system (i.e. closed
system) level, where one can easily identify the QCA between the quantum
von Neumann equation $\frac{d}{dt}\rho_{t}=-\frac{i}{\hbar}\,\left[H(t),\:\rho_{t}\right]$
and the classical Liouville equation $\frac{\partial}{\partial t}\rho_{t}(x)=\left\{ H_{t}(x),\,\rho_{t}(x)\right\} _{x}$.

The apparent tension between the classical and quantum scenarios in
the case of open system lies in the second term of Eq.(\ref{eq:eom 1st order}),
namely $B_{t}(x)\rho_{t}^{S}(x)$, which does not appear to have a
counterpart in the first-order quantum master equation in \cite{yu}.
In the followings, we will present two ways of possibly getting around
this apparent tension.
\end{singlespace}
\begin{singlespace}

\subsection*{Possible solution 1}
\end{singlespace}

\begin{singlespace}
A closer look at the prefactor $B_{t}(x)\equiv\int dx'\left\{ H_{t}(x,x'),\,\rho_{0}^{E}(x')\right\} _{x'}$
might offer some hint. For an integral over the environmental variables
$x'\equiv\left\{ q'_{1},\,p'_{1},\,q'_{2},\,p'_{2},\ldots q'_{N_{E}},\,p'_{N_{E}}\right\} $,
we have 
\begin{eqnarray}
B_{t}(x) & \equiv & \int dx'\left\{ H_{t}(x,x'),\,\rho_{0}^{E}(x')\right\} _{x'}\nonumber \\
 & = & \prod_{j=1}^{N_{E}}\int_{-\infty}^{+\infty}dq'_{j}\int_{-\infty}^{+\infty}dp'_{j}\sum_{k=1}^{N_{E}}\left(\frac{\partial H_{t}(x,x')}{\partial q'_{k}}\frac{\partial\rho_{0}^{E}(x')}{\partial p'_{k}}-\frac{\partial H_{t}(x,x')}{\partial p'_{k}}\frac{\partial\rho_{0}^{E}(x')}{\partial q'_{k}}\right)\nonumber \\
 & = & \sum_{k=1}^{N_{E}}\left(\prod_{j=1}^{N_{E}}\int_{-\infty}^{+\infty}dq'_{j}\int_{-\infty}^{+\infty}dp'_{j}\left(\frac{\partial H_{t}(x,x')}{\partial q'_{k}}\frac{\partial\rho_{0}^{E}(x')}{\partial p'_{k}}-\frac{\partial H_{t}(x,x')}{\partial p'_{k}}\frac{\partial\rho_{0}^{E}(x')}{\partial q'_{k}}\right)\right)\nonumber \\
 & = & \sum_{k=1}^{N_{E}}\{\prod_{j\neq k}\int_{-\infty}^{+\infty}dq'_{j}\int_{-\infty}^{+\infty}dp'_{j}\nonumber \\
 &  & \left(\int_{-\infty}^{+\infty}dq'_{k}\int_{-\infty}^{+\infty}dp'_{k}\left(\frac{\partial H_{t}(x,x')}{\partial q'_{k}}\frac{\partial\rho_{0}^{E}(x')}{\partial p'_{k}}-\frac{\partial H_{t}(x,x')}{\partial p'_{k}}\frac{\partial\rho_{0}^{E}(x')}{\partial q'_{k}}\right)\right)\}\nonumber \\
 & \equiv & \sum_{k=1}^{N_{E}}\left(\prod_{j\neq k}\int_{-\infty}^{+\infty}dq'_{j}\int_{-\infty}^{+\infty}dp'_{j}B_{t,k}(x,x')\right),\label{eq:B step1}
\end{eqnarray}
where in the last line we have defined the factor $B_{t,k}(x,x')$
accordingly. Let's examine the factor $B_{t,k}(x,x')$ for an arbitrary
$k$ in Eq.(\ref{eq:B step1}) in greater detail: 
\begin{eqnarray}
B_{t,k}(x,x') & \equiv & \int_{-\infty}^{+\infty}dq'_{k}\int_{-\infty}^{+\infty}dp'_{k}\left(\frac{\partial H_{t}(x,x')}{\partial q'_{k}}\frac{\partial\rho_{0}^{E}(x')}{\partial p'_{k}}-\frac{\partial H_{t}(x,x')}{\partial p'_{k}}\frac{\partial\rho_{0}^{E}(x')}{\partial q'_{k}}\right)\nonumber \\
 & = & \int_{-\infty}^{+\infty}dq'_{k}\left(\int_{-\infty}^{+\infty}dp'_{k}\frac{\partial H_{t}(x,x')}{\partial q'_{k}}\frac{\partial\rho_{0}^{E}(x')}{\partial p'_{k}}\right)\nonumber \\
 &  & -\int_{-\infty}^{+\infty}dp'_{k}\left(\int_{-\infty}^{+\infty}dq'_{k}\frac{\partial H_{t}(x,x')}{\partial p'_{k}}\frac{\partial\rho_{0}^{E}(x')}{\partial q'_{k}}\right)\nonumber \\
 & = & \int_{-\infty}^{+\infty}dq'_{k}\left(\frac{\partial H_{t}(x,x')}{\partial q'_{k}}\rho_{0}^{E}(x')\mid_{p'_{k}=-\infty}^{p'_{k}=+\infty}-\int_{-\infty}^{+\infty}dp'_{k}\frac{\partial^{2}H_{t}(x,x')}{\partial p'_{k}\partial q'_{k}}\rho_{0}^{E}(x')\right)\nonumber \\
 &  & -\int_{-\infty}^{+\infty}dp'_{k}\left(\frac{\partial H_{t}(x,x')}{\partial p'_{k}}\rho_{0}^{E}(x')\mid_{q'_{k}=-\infty}^{q'_{k}=+\infty}-\int_{-\infty}^{+\infty}dq'_{k}\frac{\partial^{2}H_{t}(x,x')}{\partial q'_{k}\partial p'_{k}}\rho_{0}^{E}(x')\right)\nonumber \\
 & = & -\int_{-\infty}^{+\infty}dq'_{k}\int_{-\infty}^{+\infty}dp'_{k}\frac{\partial^{2}H_{t}(x,x')}{\partial p'_{k}\partial q'_{k}}\rho_{0}^{E}(x')\nonumber \\
 &  & +\int_{-\infty}^{+\infty}dp'_{k}\int_{-\infty}^{+\infty}dq'_{k}\frac{\partial^{2}H_{t}(x,x')}{\partial q'_{k}\partial p'_{k}}\rho_{0}^{E}(x')\nonumber \\
 & = & 0,\label{eq:B step2}
\end{eqnarray}
where in the third equality we have used integration by parts, in
the fourth equality we have assumed the following quantities vanish
at infinity (for an arbitrary $k$),
\begin{equation}
\frac{\partial H_{t}(x,x')}{\partial q'_{k}}\rho_{0}^{E}(x')\mid_{p'_{k}=\pm\infty}=\frac{\partial H_{t}(x,x')}{\partial p'_{k}}\rho_{0}^{E}(x')\mid_{q'_{k}=\pm\infty}=0,\label{eq:condition 1}
\end{equation}
and in the last equality we have assumed the symmetry of second derivatives
(for an arbitrary $k$), that is, 
\begin{equation}
\frac{\partial^{2}H_{t}(x,x')}{\partial q'_{k}\partial p'_{k}}=\frac{\partial^{2}H_{t}(x,x')}{\partial p'_{k}\partial q'_{k}},\label{eq:condition 2}
\end{equation}
and throughout we have assumed the orders of integrals are interchangeable.
With these assumptions (under these conditions), we may say $B_{t,k}(x,x')=0$
for an arbitrary $k$; plugging this into Eq.(\ref{eq:B step1}),
we may say $B_{t}(x)=0$.

Thus we have shown that the extra term $B_{t}(x)\rho_{t}^{S}(x)$
in Eq.(\ref{eq:eom 1st order}) vanishes with the aforementioned assumptions.
One might argue that these assumptions/conditions are apparently satisfiable
in ``normal'' (i.e. sufficiently ``well-behaved'') situations.
For example, looking at Eq.(\ref{eq:condition 1}), we note that the
probability density $\rho_{0}^{E}(x')$ should vanish at infinity
because of normalization requirement for a probability distribution
$\int dx'\rho_{0}^{E}(x')=1$, thus provided that $\frac{\partial H_{t}(x,x')}{\partial q'_{k}}\mid_{p'_{k}=\pm\infty}$
and $\frac{\partial H_{t}(x,x')}{\partial p'_{k}}\mid_{q'_{k}=\pm\infty}$
do not blow up, the condition Eq.(\ref{eq:condition 1}) would be
satisfied. With the vanishing of $B_{t}(x)\rho_{t}^{S}(x)$, the first-order
equation of motion Eq.(\ref{eq:eom 1st order}) would then become
\begin{eqnarray}
\frac{\partial}{\partial t}\rho_{t}^{S}(x) & = & \left\{ H_{t}^{eff}(x),\,\rho_{t}^{S}(x)\right\} _{x},\label{eq:eom 1st order new}
\end{eqnarray}
which we might now say has an apparent QCA with the first-order quantum
master equation $\frac{d}{dt}\rho_{t}^{S}=-\frac{i}{\hbar}\,\left[H_{eff}(t),\:\rho_{t}^{S}\right]$.
It should be stressed that we do not claim rigour in these arguments.

One might still ask whether we have satisfactorily established quantum-classical
analogy with this approach, because the vanishing of $B_{t}(x)\rho_{t}^{S}(x)$
is apparently not a priori. If in the case of first-order quantum
master equation there were a priori nothing beyond the commutator
$-\frac{i}{\hbar}\,\left[H_{eff}(t),\:\rho_{t}^{S}\right]$, whereas
in the case of classical equation of motion there were something beyond
the Poisson bracket $\left\{ H_{t}^{eff}(x),\,\rho_{t}^{S}(x)\right\} _{x}$,
which turned out to vanish given some conditions (albeit possibly
``reasonable'' ones), does this adequately qualify as QCA? Or is
it too weak a sense to talk about QCA?
\end{singlespace}
\begin{singlespace}

\subsection*{Possible solution 2}
\end{singlespace}

\begin{singlespace}
In response to the questions above, let's have a closer look at the
first-order quantum master equation in \cite{yu}. For $H_{SE}(t)=\sum_{n}S_{n}(t)\otimes E_{n}(t)$,
the original expression for the first-order term is 
\begin{eqnarray}
\mathcal{L}_{1,t}\left(\rho_{t}^{S}\right) & = & -\frac{i}{\hbar}\left(\sum_{n}Tr_{E}\left(S_{n}(t)\rho_{t}^{S}\otimes E_{n}(t)\rho_{E0}\right)-\sum_{n}Tr_{E}\left(\rho_{t}^{S}S_{n}(t)\otimes\rho_{E0}E_{n}(t)\right)\right),\nonumber \\
\end{eqnarray}
and thus the first-order master equation may be worked out more carefully
as follows:
\begin{eqnarray}
\frac{d}{dt}\rho_{t}^{S} & = & -\frac{i}{\hbar}\left(\sum_{n}Tr_{E}\left(S_{n}(t)\rho_{t}^{S}\otimes E_{n}(t)\rho_{E0}\right)-\sum_{n}Tr_{E}\left(\rho_{t}^{S}S_{n}(t)\otimes\rho_{E0}E_{n}(t)\right)\right)\nonumber \\
 & = & -\frac{i}{\hbar}\sum_{n}\left(S_{n}(t)\rho_{t}^{S}Tr_{E}\left(E_{n}(t)\rho_{E0}\right)-\rho_{t}^{S}S_{n}(t)Tr_{E}\left(\rho_{E0}E_{n}(t)\right)\right)\nonumber \\
 & = & -\frac{i}{\hbar}\sum_{n}\left(S_{n}(t)\rho_{t}^{S}Tr_{E}\left(\rho_{E0}E_{n}(t)\right)-\rho_{t}^{S}S_{n}(t)Tr_{E}\left(\rho_{E0}E_{n}(t)\right)\right)\nonumber \\
 &  & -\frac{i}{\hbar}\sum_{n}\left(S_{n}(t)\rho_{t}^{S}Tr_{E}\left(E_{n}(t)\rho_{E0}\right)-S_{n}(t)\rho_{t}^{S}Tr_{E}\left(\rho_{E0}E_{n}(t)\right)\right)\nonumber \\
 & = & -\frac{i}{\hbar}\left[\sum_{n}Tr_{E}\left(\rho_{E0}E_{n}(t)\right)S_{n}(t),\,\rho_{t}^{S}\right]\nonumber \\
 &  & -\frac{i}{\hbar}\sum_{n}\left(Tr_{E}\left(S_{n}(t)\otimes E_{n}(t)\left(\mathbb{I}_{S}\otimes\rho_{E0}\right)\right)\rho_{t}^{S}-Tr_{E}\left(\left(\mathbb{I}_{S}\otimes\rho_{E0}\right)S_{n}(t)\otimes E_{n}(t)\right)\rho_{t}^{S}\right)\nonumber \\
 & = & -\frac{i}{\hbar}\left[\sum_{n}Tr_{E}\left(S_{n}(t)\otimes\rho_{E0}E_{n}(t)\right),\,\rho_{t}^{S}\right]\nonumber \\
 &  & -\frac{i}{\hbar}\left(Tr_{E}\left(\left(\sum_{n}S_{n}(t)\otimes E_{n}(t)\right)\left(\mathbb{I}_{S}\otimes\rho_{E0}\right)\right)\rho_{t}^{S}\right)\nonumber \\
 &  & -\frac{i}{\hbar}\left(-Tr_{E}\left(\left(\mathbb{I}_{S}\otimes\rho_{E0}\right)\left(\sum_{n}S_{n}(t)\otimes E_{n}(t)\right)\right)\rho_{t}^{S}\right)\nonumber \\
 & = & -\frac{i}{\hbar}\left[Tr_{E}\left(\left(\mathbb{I}_{S}\otimes\rho_{E0}\right)\sum_{n}S_{n}(t)\otimes E_{n}(t)\right),\,\rho_{t}^{S}\right]\nonumber \\
 &  & -\frac{i}{\hbar}\left(Tr_{E}\left(H_{SE}(t)\left(\mathbb{I}_{S}\otimes\rho_{E0}\right)\right)\rho_{t}^{S}-Tr_{E}\left(\left(\mathbb{I}_{S}\otimes\rho_{E0}\right)H_{SE}(t)\right)\rho_{t}^{S}\right)\nonumber \\
 & = & -\frac{i}{\hbar}\left[Tr_{E}\left(\left(\mathbb{I}_{S}\otimes\rho_{E0}\right)H_{SE}(t)\right),\,\rho_{t}^{S}\right]\nonumber \\
 &  & +Tr_{E}\left(-\frac{i}{\hbar}\,\left[H_{SE}(t),\,\left(\mathbb{I}_{S}\otimes\rho_{E0}\right)\right]\right)\rho_{t}^{S}\nonumber \\
 & = & -\frac{i}{\hbar}\left[H_{eff}(t),\,\rho_{t}^{S}\right]+B(t)\rho_{t}^{S},\label{eq:L1}
\end{eqnarray}
where 
\begin{eqnarray}
H_{eff}(t) & \equiv & Tr_{E}\left(\left(\mathbb{I}_{S}\otimes\rho_{E0}\right)H_{SE}(t)\right),\\
B(t) & \equiv & Tr_{E}\left(-\frac{i}{\hbar}\,\left[H_{SE}(t),\,\left(\mathbb{I}_{S}\otimes\rho_{E0}\right)\right]\right).
\end{eqnarray}
Now we see an extra term $B(t)\rho_{t}^{S}$ beyond the commutator
$-\frac{i}{\hbar}\left[H_{eff}(t),\,\rho_{t}^{S}\right]$ in the first-order
quantum master equation. Note that the difference between this derivation
and the original derivation in \cite{yu}, the latter of which does
not gives rise to the term $B(t)\rho_{t}^{S}$, lies in the third
equality of Eq.(\ref{eq:L1}). In the original derivation in \cite{yu}
we simply assume $Tr_{E}\left(\rho_{E0}E_{n}(t)\right)=Tr_{E}\left(E_{n}(t)\rho_{E0}\right)$,
whereas here in Eq.(\ref{eq:L1}) we do not assume the same, but formally
work out the difference between $Tr_{E}\left(\rho_{E0}E_{n}(t)\right)$
and $Tr_{E}\left(E_{n}(t)\rho_{E0}\right)$. The result of this new
treatment is the second term of Eq.(\ref{eq:L1}), namely $B(t)\rho_{t}^{S}$.

Comparing the first-order quantum master equation Eq.(\ref{eq:L1})
with its classical counterpart Eq.(\ref{eq:eom 1st order}), we may
now see a better analogy between them. In particular, the second term
of Eq.(\ref{eq:eom 1st order}) 
\begin{equation}
B_{t}(x)\rho_{t}^{S}(x)=\int dx'\left\{ H_{t}(x,x'),\,\rho_{0}^{E}(x')\right\} _{x'}\rho_{t}^{S}(x)
\end{equation}
 now has a analogous term in Eq.(\ref{eq:L1}) 
\begin{equation}
B(t)\rho_{t}^{S}=Tr_{E}\left(-\frac{i}{\hbar}\,\left[H_{SE}(t),\,\left(\mathbb{I}_{S}\otimes\rho_{E0}\right)\right]\right)\rho_{t}^{S},
\end{equation}
if we were again to identify $\left\{ A(x),\,B(x)\right\} _{x'}$
and $\int dx'\left(\ldots\right)$ in the former with $-\frac{i}{\hbar}\,\left[A,\:B\right]$
and $Tr_{E}\left(\ldots\right)$ in the latter. In view of this, one
may argue that QCA is plausibly re-established.

One might still ask further questions, for example, whether the vanishing
of $B(t)\rho_{t}^{S}$ in the quantum case is subject to analogous
(or ``corresponding'') conditions as the vanishing of $B_{t}(x)\rho_{t}^{S}(x)$
in the classical case is, so that whenever one vanishes the other
would also vanish accordingly/correspondingly.

Regardless of the possibility of further questions, the formal analogy
between $B_{t}(x)\rho_{t}^{S}(x)$ in the classical case and $B(t)\rho_{t}^{S}$
in the quantum case apparently holds anyway. One might even argue
that, in some sense, this might be the right kind of quantum-classical
analogy one would be looking for.
\end{singlespace}
\begin{singlespace}

\subsection*{Summary}
\end{singlespace}

\begin{singlespace}
In the above, we might have possibly/arguably established quantum-classical
analogy at the level of open system dynamics, up to the leading (first)
perturbative order. It should be noted that we do not have a definitive
conclusion regarding whether QCA is sufficiently established here,
nor do we claim rigour in our discussions. We only mean to provide
some heuristics for the issue and possibly some starting point for
further discussions.

Besides the issue of QCA per se, this work also helps us notice the
formal existence of the term $B(t)\rho_{t}^{S}$ in the first-order
quantum master equation, an issue that was neglected in \cite{yu}.
Thus one contribution our formalism could have is that the classical
equation of motion derived herein might shed light on its quantum
counterpart and possibly help us notice issues that might have been
overlooked in the works of open quantum systems.
\end{singlespace}
\begin{singlespace}

\section{Conclusions }
\end{singlespace}

\begin{singlespace}
In this work we develop a formalism to study the evolution of the
marginal probability density of open classical systems, using Green's
functions and a series expansion method.

We have worked out the equation of motion for a general open classical
system's marginal probability density up to first order, namely Eq.(\ref{eq:eom 1st order}).
With this framework, higher-order terms can be worked out systematically,
by using the $Y_{N,t}$ map as in Eq.(\ref{eq:Y map}). (Also, see
\cite{yu} for a treatment using the $Y_{N,t}$ map in the context
of open quantum systems.) By working out higher-order terms, one will
be able to see how system-environment coupling further modifies the
(idealistically Liouville) dynamics of classical systems. For example,
it may be interesting to examine the second-order term. In the case
of open quantum systems, quantum decoherence generally arises as a
second-order effect. \cite{yu} It might be interesting if we could
see some counterpart of ``decoherence'' in the context of open classical
system's statistics using our formalism.

Besides such general study of a formal nature, our formalism may be
applied to real-world classical systems to study the evolution of
their (marginal) statistics when coupled to external degrees of freedom.
This practicality can be a direction for future efforts.

This work may also contribute to the study of quantum-classical analogy
\cite{dragoman book} from an open system dynamics perspective, by
comparing the first-order equation of motion Eq.(\ref{eq:eom 1st order})
(or any higher-order equation of motion that can be worked out with
our framework) with what could be considered its quantum counterpart
(e.g. the quantum master equation in \cite{yu}). We notice an apparent
tension between quantum and classical open system dynamics in first
order; we also present possible ways of getting around the tension.
However, we claim no definitive conclusion about a sufficient quantum-classical
analogy, but only hope to provide some starting point for further
investigations along this line.
\end{singlespace}

\end{document}